\documentclass[preprint,groupedaddress]{revtex4-1}

\usepackage{color}
\usepackage{graphicx}
\usepackage{epstopdf}
\usepackage{soul}
\usepackage{amsmath,amssymb}
\usepackage{threeparttable}

\begin{document}

\title{An integrated quantum photonic sensor based on Hong-Ou-Mandel interference}

\author{Sahar Basiri-Esfahani$^1$\footnote{s.basiri@uq.edu.au}, Casey R. Myers$^1$, Ardalan Armin$^2$, Joshua Combes$^{1,3}$ and Gerard J. Milburn$^1$}
\affiliation{$^1$ARC Centre for Engineered Quantum Systems, School of Mathematics and Physics, The University of Queensland, St Lucia, QLD 4072, Australia}
\affiliation{$^2$Centre for Organic Photonics \& Electronics (COPE), School of Mathematics and Physics and School of Chemistry and Molecular Biosciences, The University of Queensland,
St Lucia, QLD 4072, Australia}
\affiliation{$^3$Center for quantum information and control, University of New Mexico, Albuquerque, New Mexico 87131-0001, USA}

\begin{abstract}
Photonic-crystal-based integrated optical systems have been used for a broad range of sensing applications with great success.  This has been motivated by several advantages such as high sensitivity, miniaturization, remote sensing, selectivity and stability. Many photonic crystal sensors have been proposed with various fabrication designs that result in improved optical properties. In parallel, integrated optical systems are being pursued as a platform for photonic quantum information processing using linear optics and Fock states. Here we propose a novel integrated Fock state optical sensor architecture that can be used for force, refractive index and possibly local temperature detection. In this scheme, two coupled cavities behave as an ``effective beam splitter". The sensor works based on fourth order interference (the Hong-Ou-Mandel effect) and requires a sequence of single photon pulses and consequently has low pulse power. Changes in the parameter to be measured induce variations in the effective beam splitter reflectivity and result in changes to the visibility of interference. We demonstrate this generic scheme in coupled L3 photonic crystal cavities as an example and find that this system, which only relies on photon coincidence detection and does not need any spectral resolution, can estimate forces as small as $10^{-7}$ Newtons and can measure one part per million change in refractive index using a very low input power of $10^{-10}$W.  Thus linear optical quantum photonic architectures can achieve  comparable sensor performance to semiclassical devices. 
\end{abstract}

\maketitle

\section{Introduction}
Integrated photonics based on photonic crystal (PhC) structures provides a path to extremely small optical sensors with applications to biology~\cite{Heeres,Scullion}, chemistry~\cite{Zhang1} and engineering~\cite{Pinto}. PhC devices with various geometries and structures such as hollow core PhC fibers~\cite{Frazao}, 1D and 2D waveguides~\cite{Hopman,Buswell} and nano-cavities~\cite{data2,HaddadiAPL,Atlasov,L3_4,data} have been fabricated and used for sensing applications. Among these devices, PhC cavity-based sensors offer important advantages over PhC waveguide sensors since they can be made much smaller, thus reducing vulnerability from impurities and losses. Moreover, exploiting high Q cavities with large mode volume are advantageous for sensors based on refractive index (RI) changes, for example in bio-pathogen detection~\cite{Chakravarty}, chemical sensing~\cite{Falco} and single particle detection~\cite{Lee1}. All these schemes make use of second order interference of coherent states of the optical field and are thus basically classical phenomenon.  

In parallel to integrated optical sensors, considerable progress has been in using integrated optical systems for single photon optical quantum computing using linear optics, so called LOQC~\cite{nature-Gerard}. These schemes are enabled by the uniquely quantum-optical phenomenon of Hong-Ou-Mandel (HOM) interference~\cite{HOM}. This opens up a new perspective for optical quantum metrology that combines ideas from photonic crystal sensors with linear optical quantum information processing using single photon states.   We make a first step in this direction by presenting a scheme  that uses HOM interference to make sensors from optical pulses prepared in single photon pulses. As HOM interference does not arise for coherent optical pulses, our proposal is a true quantum metrology scheme and realizes the gain in sensitivity such schemes offer. Furthermore, it opens up the prospect of using LOQC protocols to construct more sophisticated quantum metrology protocols that are compatible with integrated optical systems.  

Recent demonstrations of cutting edge sensors that exploit quantum mechanics have been shown to outperform their classical counterparts in achieving higher sensitivities~\cite{Lukin,Kippenberg,Wineland}. Many applications, e.g. biological sensing~\cite{Bowen},  require low power to preserve delicate samples destroyed by: photo-decomposition, photo-thermal effects, and photon pressure for example. This requirement is in addition to the usual requirements of high input-output gain (responsivity), low noise and high bandwidth. In that regard, weak coherent light offers a route to low power sensing. However, the use of weak coherent pulses lowers a sensor's bandwidth.  Consider for example a series of weak coherent pulses with on average one photon per pulse, in this case roughly $37\%$ of pulses have no photons at all and $26\%$ have more than one photon per pulse. Clearly the ultimate low pulse power limit is achieved by single photon pulses with only one photon per pulse. A sensor operating with single photon states offers low power suitable for deployment in {\em lab on a chip} applications~\cite{Walmsley} and compatible with attojoule all-optical switching~\cite{LP-switch-nphot} and opto-mechanical devices for strain sensors~\cite{Painter2013} and accelerometers~\cite{Painter2012}.

While single photon states are not easy to make there is a very large research effort underway driven by their potential application in quantum information processing~\cite{nature-Gerard}. For our purposes it suffices to note that PhC devices are compatible with a number of quantum dot single photon sources~\cite{opt.comm.} and that technological advances in integrated multiplexed single photon sources in PhCs are very encouraging~\cite{Collins}. The fundamental quantum nature of photons is usually observed through the HOM effect which has now been demonstrated in a variety of physical systems such as evanescently coupled optical waveguides~\cite{Politi} and microwave devices~\cite{Lang}. In the HOM effect indistinguishable photons simultaneously arrive at each of the two input ports of a 50/50 beam splitter, after which the photons ``bunch'' together so that both photons are either in one output port or the other. Never will you observe one photon in both outputs.

\begin{figure}
\centering
\includegraphics[width=0.7\linewidth]{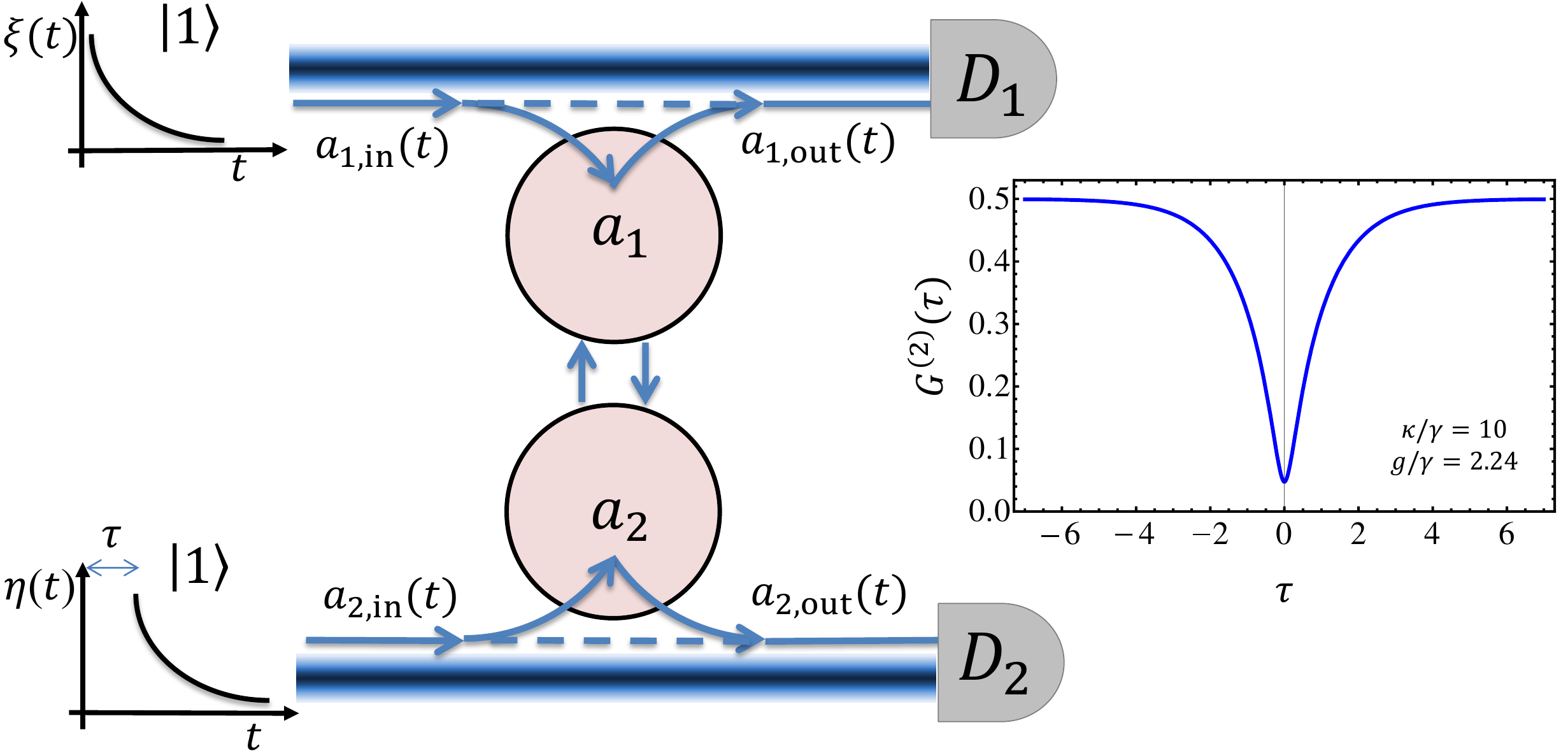}
\caption{Schematic of quantum PhC sensor. Coupled PhC resonators implement an effective beam splitter interaction between the input single photon fields resulting in HOM interference effect observed in detected output fields. $G^{(2)}(\tau)$ is a measure of the number of coincidences which is a function of the time shift between the photons entering the beam splitter. By compressing or stretching the distance between optical resonators or through changes in refractive index of the medium between the resonators, the coupling between the cavities changes. This results in a change in transmission and reflection of the beam splitter and therefore results in a change in the measured HOM visibility. }
\label{fig1}
\end{figure}

The use of dual Fock states was proposed in 1993 in the context of quantum metrology to reduce the uncertainty of phase measurements~\cite{Holland}. In this article, we propose a novel scheme for a quantum photonic sensor based on coupled PhC cavities that exploits the HOM effect, shown in Fig. \ref{fig1}. The coupled PhC cavities form an ``effective beam splitter" for two incident photons. The central idea is that a parameter to be estimated, call  it $\psi$, modulates the coupling between the optical cavities, $g$. This can be done by changing the distance between the cavities through compressing or stretching the dielectric material (e.g. for force and strain sensing) or by changing the refractive index of the media between the two cavities (e.g. for RI, temperature and single particle sensing). The change in $g$ modifies the reflection and transmission of the effective beam splitter which changes the visibility of HOM interference. Therefore, by measuring the change in HOM visibility, we can sense the variation in $g$ and thus estimate $\psi$. This scheme is independent of transmission/reflection spectra normally used for classical cavity-based sensors~\cite{Chow} and neither a dispersive element nor spectral resolution for the measurement is required. 

First, we characterize the proposed sensor in terms of its performance metrics, the responsivity and minimum detectable value for the parameter to be estimated. This characterization in terms of the working parameters of the sensor is expressed in a general way, with no assumptions set for the values for the cavity damping rate, cavities coupling strength, PhC refractive index, etc. Then, a more specific example is provided by considering this scheme with previously reported experimental parameters for GaAs/AlGaAs PhC structures. We theoretically predict that such a system can measure one part per million change in refractive index as well as forces on the order of $10^{-7}$ N. These results are not obtained by using experimental values specifically optimized for our scheme. However, the results obtained for refractive index and force sensing are promising for integrated on-chip sensing.

\section{Hong-Ou-Mandel sensor}
\label{Hong-Ou-Mandel sensor}

As HOM interference is a uniquely quantum mechanical phenomenon we must necessarily proceed with a full quantum description. Consider the double optical resonator scheme composed of two optical cavities with resonance frequency $\omega$, depicted in Fig. \ref{fig1}. The optical fields in the two cavity modes are described by the bosonic annihilation operators $a_1$ and $a_2$. The interaction picture Hamiltonian is given by
\begin{equation}
 H_I= \hbar g(a_1^\dagger a_2+a_1 a_2^\dagger),
 \label{Hamiltonian}
\end{equation}
where $g$ is the effective interaction strength that depends on the parameter $\psi$. We couple this system, via the evanescent field, to the input and output channels comprising two optical wave-guides. The relation between the respective input and output fields is \cite{Qnoise}
\begin{equation}
a_{j,{\rm out}}(t) =\sqrt{\kappa_j}a_j(t)-a_{j,{\rm in}}(t),
\label{in-out}
\end{equation}
where $\kappa_j (j=1,2)$ is the damping rate of cavity $j$. Cavity modes $a_1(t)$ and $a_2(t)$ can be related to the input modes using the input-output stochastic differential equations \cite{Qnoise,QObook}  
\begin{eqnarray}
 \frac{da_1(t)}{dt}=-ig a_2(t)-\frac{\kappa}{2}a_1(t)+\sqrt{\kappa}a_{1,{\rm in}}(t) , \nonumber \\
 \frac{da_2(t)}{dt}=-ig a_1(t)-\frac{\kappa}{2}a_2(t)+\sqrt{\kappa}a_{2, {\rm in}}(t), 
 \label{stochastic_eqns}
 \end{eqnarray}
where the solution to these equations is given in appendix A. To operate this device as a sensor we then load the two input ports with single photons and perform coincidence detection at the outputs. The upper input port of the beam splitter is loaded with a single photon in the state $\vert1_{\xi}\rangle=\int ds\xi(s)a_{1,{\rm in}}^{\dagger}(s)\vert0\rangle$ having pulse shape $\xi(t)=\sqrt{\gamma} e^{-\frac{1}{2}\gamma t}$ with the normalization condition $\int dt\vert\xi(t)\vert^2=1$. The lower input port is loaded with a single photon having exactly the same amplitude function but time shifted with respect to the top $\eta(t)=\xi(t-\tau)=\sqrt{\gamma} e^{-\frac{1}{2}\gamma (t-\tau)}$, where $\gamma$ is the input photon bandwidth. This state has zero field amplitude, $\langle a_{{\rm in}}(t)\rangle =0$, so conventional (second order) interference cannot be used. However $\langle a_{{\rm in}}^\dagger(t)a_{{\rm in}}(t)\rangle \neq 0$ so fourth order interference will reflect the quantum coherence inherent in the pure state $\vert1_{\xi}\rangle$.

The probability of one and only one click occurring at both detectors $D_1$ and $D_2$ is given by the fourth order correlation function
\begin{equation}
G^{(2)}(\tau)= \frac{\int_0^\infty \int_0^\infty \langle a_{1,{\rm out}}^\dagger (t)a_{2,{\rm out}}^\dagger (t')a_{2,{\rm out}}(t')a_{1,{\rm out}}(t)\rangle dtdt'}{\int_0^\infty\langle a_{1,{\rm out}}^\dagger (t) a_{1,{\rm out}}(t)\rangle dt\int_0^\infty\langle a_{2,{\rm {\rm out}}}^\dagger (t) a_{2,{\rm out}}(t)\rangle dt}.
\label{Pc}
\end{equation}
It should be noted that in this expression the time $\tau$ is not the delay between detection events but a temporal separation of the two input photons. In practice, the integration time need not and should not be infinite as it sets the time interval between successive pulses. In fact the integration time needs to be of the order $\tau_{\rm rep}\sim \max \{ 1/\kappa, 1/\gamma \}$. In what follows we work in regimes where $\dfrac{\kappa}{\gamma}>1$, which is compatible with available experimental realizations, so we have $\tau_{\rm rep}\sim 1/\gamma$. Through equations (\ref{Hamiltonian}), (\ref{in-out}) and (\ref{stochastic_eqns}) the explicit dependence of $G^{(2)}(\tau)$ on $g$ can be seen. By monitoring changes in $G^{(2)}(\tau)$ we can infer changes in $g$. In the ideal case, we would like to detect both photons. 

In practice, either one or two photons could fail to be counted at the detectors due to optical losses, imperfect single photon sources or non-unit detection efficiency. The case of photon loss before detection (or detector inefficiency) can be modelled as usual~\cite{loss} by inserting a beam splitter with transmissivity amplitude $t_k$ in the path of the output fields $a_{k,{\rm out}}(t)$. The field that reaches an ideal detector is given by the transformation $a_{1,{\rm out}}(t)\rightarrow  t_k a_{1,{\rm out}}(t)+r_k v_{1,{\rm out}}(t)$ where $v(t)$ is a vacuum field mode annihilation operator. Substitution into the equation (\ref{Pc}) shows that the only term that contributes to the both the numerator and denominator are  multiplied by the factor $T_1.T_2$, where $T_k=|t_k|^2$ is the conditional probability for a single photon in the output field to reach an ideal detector. This is because this average is normally ordered. Thus $G^{(2)}(\tau)$ is unchanged by loss since we have normalised it by the intensity that actually reaches the detector from each mode: in effect $G^{(2)}(\tau)$ is a conditional probability conditioned on only those detection events that give two counts, one at each detector. Single counts and no counts are discarded. These cases should be considered as failed, but heralded trials in which we discard and simply run again with another two single photons. However, this lowers the sensor's bandwidth. In section \ref{Noise characteristics}, we explicitly include an optical loss factor, $\varepsilon$, which is defined as the number of failed trials over the total number of trials, to take the effect of these imperfections into account. As photon loss is heralded, LOQC error correction techniques might be employed to mitigate the loss of signal on such events.

Photon loss before the device, or failure of a source to produce a photon can likewise be modelled by inserting a beam splitter into the path of $a_{{\rm in}}(t)$. This changes the input state to a mixed state as follows. The input state is a two photon state of the form $a^\dagger_{1_{\rm in}}(t)a^\dagger_{2_{\rm in}}(t)|0\rangle$. We then transform $a_{1,{\rm in}}(t)\rightarrow \bar{a}_{1,{\rm in}}(t)= t_k a_{1,{\rm in}}(t)+r_k v_{1,{\rm in}}(t)$ where $v(t)$ is a vacuum field mode annihilation operator. Thus the total state after the beam splitter is given by $\bar{a}^\dagger_{1,{\rm in}}(t)\bar{a}^\dagger_{2,{\rm in}}(t)|0\rangle$ but the actual input state to the device is given by tracing out over the two vacuum modes. This gives the input state as a mixed state of the form
\begin{eqnarray}
\rho_{\rm in} & = & T_1T_2 |1\rangle_1\langle 1|\otimes|1\rangle_2\langle 1|+T_1R_2 |1\rangle_1\langle 1|\otimes|0\rangle_2\langle 0|\\\nonumber
&&+R_1T_2 |0\rangle_1\langle 0|\otimes|1\rangle_2\langle 1|+R_1R_2 |0\rangle_1\langle 0|\otimes|0\rangle_2\langle 0\vert.
\end{eqnarray}
This also indicates that loss at the input is detected as the conditional input state, conditioned on counting two photons in total at ideal detectors. This is simply the first term in the above sum. This is the same pure state as for the case of perfect sources.  The coefficient $T_1T_2$ is simply the conditional probability that two input photons in each of the inputs enter the device. Input loss or source inefficiency is also heralded in the detectors and those trials can be discarded. The fraction discarded in total including input inefficiency and detectors inefficiency is simply $T_{1,{\rm in}}T_{2,{\rm in}} T_{1,{\rm out}}T_{2,{\rm out}}$, where $T_{k,{\rm in}}$ and $T_{k,{\rm out}}$ are the conditional probabilities that a single photon enters the device in mode-k and that a single photon for output mode-k is detected. 

In Fig. \ref{fig1}, the HOM dip for our system is depicted for particular values of $\kappa/\gamma$ and $g/\gamma$. For $\tau=0$, where input photons are indistinguishable, quantum interference results in photon bunching, or photon pairs, and we see the minimum of the coincidence probability i.e. the HOM dip. As $\tau$ increases or decreases the coincidence probability increases. 
 
We define the responsivity of the sensor to detect the changes in $g$ as  
\begin{equation}
R_g(g_0,\kappa)=\left \vert\frac{dG^{(2)}(0)}{dg}\right \vert.
\end{equation}
\begin{figure}
\centering
\includegraphics[width=\linewidth]{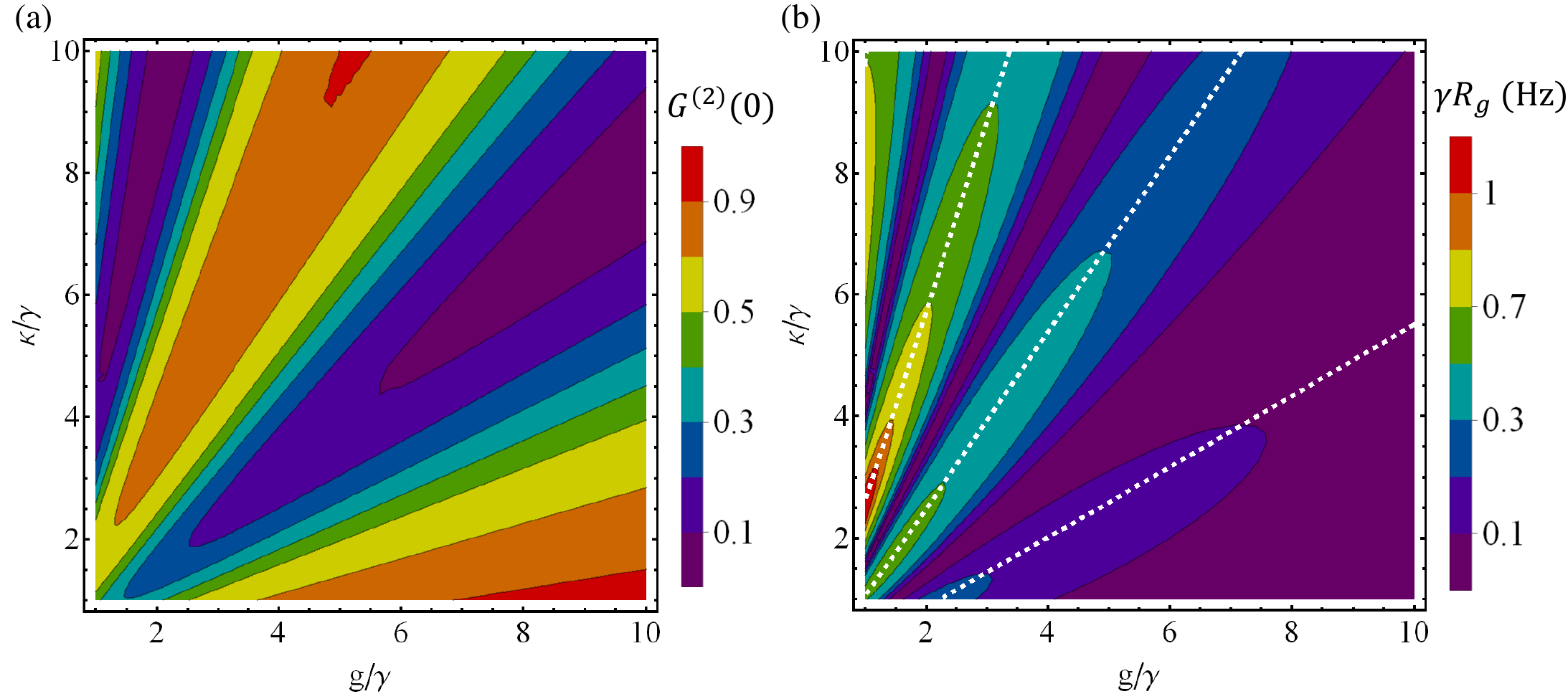}
\caption{Sensor response to variations in g. (a) Shows the behavior of the estimator, coincidence detection probability $G^{(2)}(0)$, for indistinguishable input photons versus $g/\gamma$ and $\kappa/\gamma$. (b) Shows how responsivity of the sensor varies by operating the sensor at different regimes of $g/\gamma$ and $\kappa/\gamma$. The white dashed lines show the operating points for which sensor response is maximum and linear over the range of small changes in signal.}
\label{fig2}
\end{figure}

Operating at $\tau=0$ is optimal for most combinations of $\gamma$ and $\kappa$ and maximizes the responsivity. We then optimize the values of $\kappa/\gamma$ and $g/\gamma$ so that the derivative of $G^{(2)}(0)$ with respect to $g$ is maximized. By maximizing the responsivity over our device parameters, $g_0$ the initial beam splitter coefficient and the cavity damping rate $\kappa$, we can optimize the performance of our sensor. Due to the fact that our sensor is a linear quantum system, we can analytically calculate $G^{(2)}(\tau)$ and its derivative for the initial state $\vert\psi(0)\rangle=\vert1_{a_1,\xi},1_{a_2,\eta}\rangle$, the full expression for $G^2(\tau)$ is given in appendix A. Figure \ref{fig2} can serve as a guide for experimental implementations and device fabrication. Figure \ref{fig2}(a) shows $G^{(2)}(0)$ as a function of $g/\gamma$ and $\kappa/\gamma$. Figure \ref{fig2}(b) shows the behaviour of the system response for different operating points $g/\gamma$ and $\kappa/\gamma$. The dashed line on Fig. \ref{fig2}(b) demonstrates the operating points at which $\dfrac{d R_g}{dg}=0$ where we can take advantage of maximum sensor response. In addition, at this maximum sensor response, the estimator $G^2(0)$ behaves linearly with small signal variations as will be described below. \\

\section{Noise characteristics}
\label{Noise characteristics}
 
Another important measure in characterizing the sensor performance is the Linear Dynamic Range (LDR) which is related to the estimation error and the sensor linearity which we now explore. The error in estimating $\delta g$ is related to the error in estimating $\delta G^{(2)}(0)$ in a finite number of samples
\begin{equation}
\delta g_{\rm noise}=\left |\frac{dG^{(2)}(0)}{dg}\right |^{-1}\delta G^{(2)}(0)_{\rm noise},
\label{gnoise}
\end{equation}
where 
\begin{equation}
\label{dG-noise}
\delta G^{(2)}(0)_{\rm noise}=\frac{\sqrt{G^{(2)}(0)(1-G^{(2)}(0))}}{\sqrt{N(1-\varepsilon)}},
\end{equation}
\begin{figure}
\centering
\includegraphics[width=0.7\linewidth]{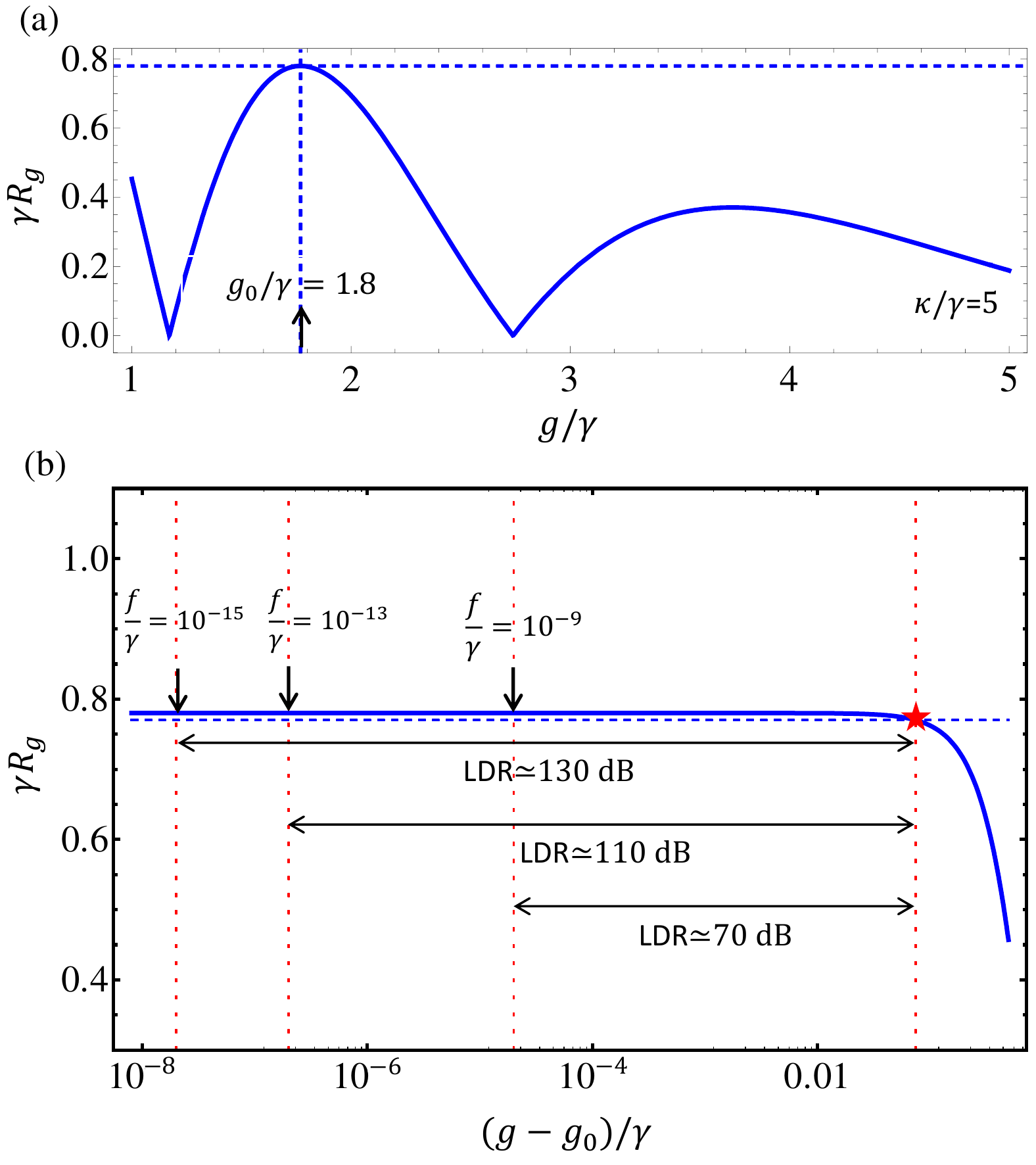}
\caption{Linear dynamic range. (a) Shows how sensor response changes at different operating points. If we operate the sensor on a bias $g_0$ where sensor response is maximum, we can take advantage of the sensor linear response, up to small variations in $g$. (b) Shows LDR for bias $\frac{g_0}{\gamma}=1.8$ shown in (a) for different detection frequency bandwidths over $\gamma$, $\frac{f}{\gamma}$. The red star shows the upper LDR limit that is the point up to which sensor responds linearly within $1\%$ variation.}
\label{fig3}
\end{figure}
is the standard deviation of a Bernoulli distribution with $N$ trials. The loss factor, $\varepsilon$, has already been introduced in section \ref{Hong-Ou-Mandel sensor}. The minimum detectable shift in $g$ from the bias $g_0$ should be larger than this error, i.e. $\delta g_{\rm min}>\delta g_{\rm noise}$, so that we are able to measure it. For a large number of samples ($N\rightarrow \infty$), $\delta g_{\rm noise}$ is negligible (up to accidental coincidences caused by dark counts or stray light). This result is useful for the estimation of a static or quasi static parameter. 

We now give an order of magnitude estimate for $\delta g_{\rm min}$ when the parameter is time varying. If $T_{\rm meas}= \tau_{\rm rep}N$ is the time between our samples of $g(t)$, naive arguments from the Nyquist-Shannon sampling theorem imply that we can not determine frequency components of $g(t)$ greater than $f=1/(2T_{\rm meas})$, which is called the detection frequency bandwidth. For a one-sigma level of confidence we should have $N\geq{\rm min}\lbrace\gamma,\kappa\rbrace/(2f)$ and the noise equivalent $\delta g$, given in equation (\ref{gnoise}), becomes
\begin{equation}
\delta g_{\rm min}>\frac{\sqrt{2fG^{(2)}(0)(1-G^{(2)}(0))}}{R_g\sqrt{{\rm min}\lbrace\gamma,\kappa\rbrace(1-\varepsilon)}}.
\end{equation}
Now we can calculate the LDR which is defined as 
\begin{equation}
\text{LDR}=20 \log \frac{\delta g_{\rm max}}{\delta g_{\rm min}},
\end{equation}
where $\delta g_{\rm max}$ is the point bellow which the sensor response is linear within $1\%$ variation, i.e. $R_g(g)= R_g^{\rm max}-0.01 R_g^{\rm max}$ in which $R_g^{\rm \max}\equiv \max\limits _{g_0, \kappa}R_g(g_0,\kappa)$. 

In Fig. \ref{fig3}(a), responsivity is plotted with respect to $g$ for an arbitrary value of $\kappa$. We bias the initial coupling between the optical resonators ($g_0$) where the responsivity peaks. Therefore, there is a range of $\delta g=g-g_0$ for which the sensor behaves linearly. LDR is shown in Fig. \ref{fig3}(b) for some arbitrary detection bandwidth in units of $\gamma$. For smaller choices of $f/\gamma$, $\delta g_{\rm noise}$ will be decreased, so the sensor can resolve smaller shifts in $g$.

\section{HOM sensor implementation}

We now consider specific physical applications for our sensor, first as a force sensor and then as a refractive index sensor, employing coupled L3 PhC cavities \cite{data2,HaddadiAPL,Atlasov,L3_4,data} experimental data to estimate its responsivity and minimum resolvable shift in signal for each case. By examining the normal mode splitting reported in these references we infer the coupling strength $g$ between the PhC resonators is of the order of $10^{11}-10^{15}$ Hz. The evanescent coupling strength between the resonators and waveguides $\kappa$ can be tailored, so that  $\kappa\sim g$ for example. Operating as a force sensor, the measured signal is the shift in cavity separation induced by an applied force or a strain, while operating as a refractive index sensor, the signal to be measured is a change in refractive index induced by the presence of a molecule dropped on the air holes between the PhC resonators, for a constant bias cavity separation. A shift in either cavity separation, call it $x$, or refractive index, call it $n$, modifies the coupling strength between the resonators which will be detected by measuring $G^2(0)$. Therefore, to give an order of magnitude estimation of the responsivity and minimum detectable signal in each case we need to investigate the dependence of $g$ as a function of $x$ and $n$. To do this we used a 1D model analysed by the  transfer matrix method \cite{Pettersson} (see Appendix B) to investigate the dependence of the cavity normal mode splitting on the change in cavity separation or refractive index. 

\begin{figure}
\centering
\includegraphics[width=\linewidth]{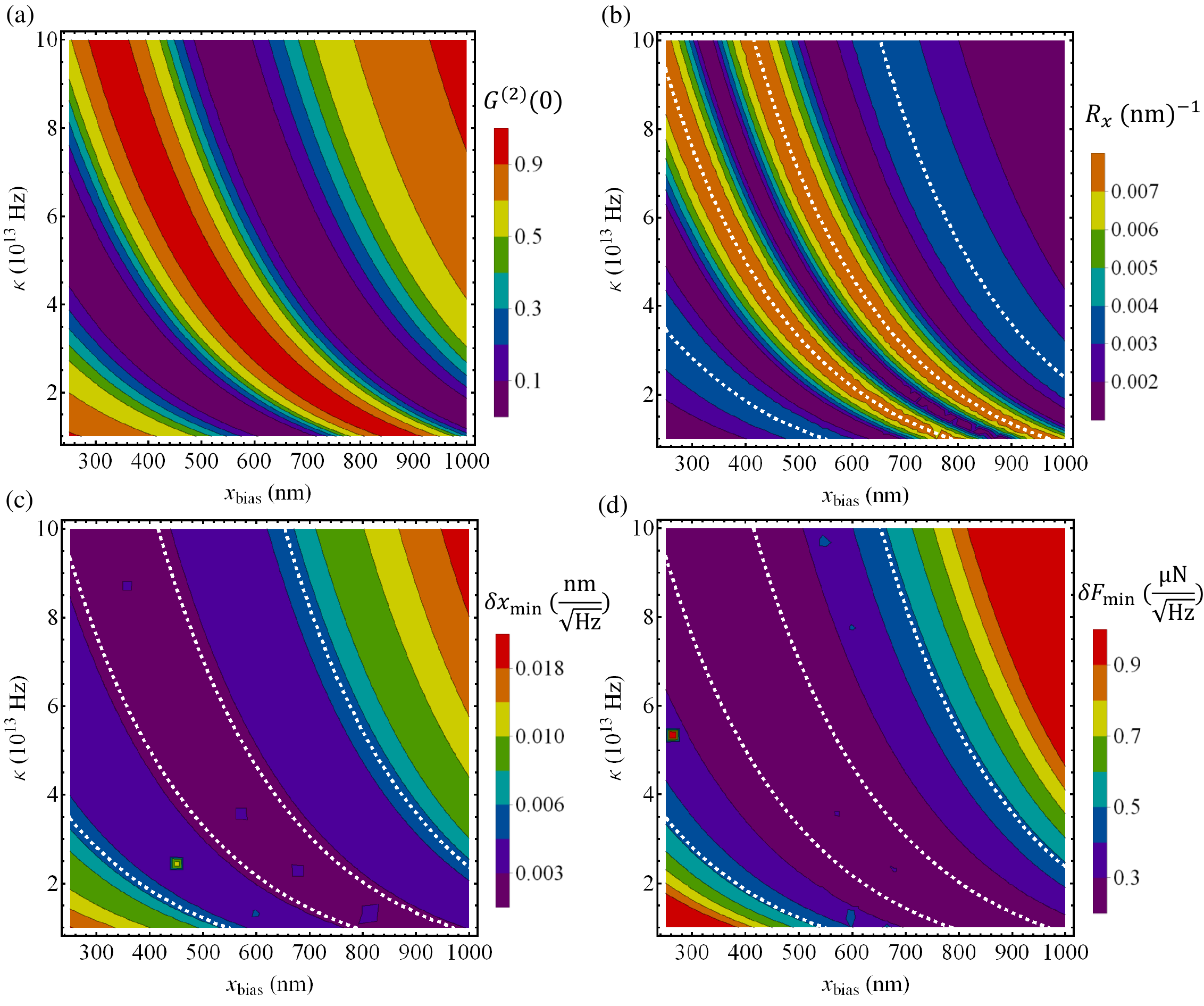}
\caption{Performance of HOM sensor as a force sensor. This figure is a fabrication guide to building a HOM force sensor with maximum performance. (a) Shows how the estimator evolves by changing the operating point $x_{\rm bias}$ for an input photon bandwidth of $\gamma=1\rm GHz$. (b) For the given $\gamma$ and $\kappa$, the responsivity is of the order of $10^{-3} (\text {nm})^{-1}$. The white dashed lines show the operating points where system response to displacement shift is linear. (c) Shows that the minimum detectable change in distance is of the order of $10^{-3}\rm nm/\sqrt{Hz})$. (d) For PhC made of GaAs/AlGaAs, the given value for minimum detectable x corresponds to minimum detectable forces of the order of $10^{-7}\rm N$. Our calculations show that as we reduce kappa, gradually we loose the linear behaviour of the sensor (white dashed lines) for smaller $x_{\rm bias}$ as the best bias points shifts towards larger $x$ or smaller g without improving or decreasing the sensor resolution. }
\label{fig4}
\end{figure}

In the case of identical resonators, $\omega_1=\omega_2=\omega$ and $\kappa_1=\kappa_2$, the splitting in frequencies of the symmetric and asymmetric normal cavity modes is $\Delta\Omega=2g$ \cite{data2,Cohen}. Therefore, we can write $g=\pi c\Delta\lambda/\lambda^2$, where $c$ is the speed of light and $\lambda$ is the cavity mode wavelength. Since $\pi c/\lambda^2$ is a constant, to find the functionality of $g$ with $x$ and $n$, we need to find the functionality of $\Delta\lambda$ with those parameters. Numerics show that an exponential function of the form $g=ae^{-bx}$ fits very well on data achieved for normal mode splitting change versus different cavity separations (see Appendix B) and an exponential of the form $g=ae^{bn^2}$ can describe the changes with respect to refractive index (see Appendix B). Hence, we can generally write $g(x,n)=ae^{-bx+dn^2}$. We extract the coefficients $a,b$ and $d$ by fitting data from figure 2 of citation \cite{data} for a PhC made of GaAs/AlGaAs (see Appendix C). According to their data $g$ is on the order of $10^{12}-10^{13}$ Hz for this range of $x_{\rm bias}$ that is shown in Fig. \ref{fig4}. We have chosen $\kappa$ of the order of $10^{13}$ Hz. 

\subsection{HOM force sensor}

First we investigate the efficiency of our system operating as a force sensor. In this case $n=1$, so by substituting $g(x,1)$ into equation (\ref{Pc}) we can see how the probability of joint detections changes for different operating points $x_{\rm bias}$ (Fig. \ref{fig4}(a)). The sensor response to changes in $x$ is calculated as $R_x(x_0,\kappa)=\vert\frac{dG^2(0)}{dx}\vert_{x=x_0}$. Figure \ref{fig4}(b) shows that for an input photon bandwidth on the order of $\gamma=1\text{GHz}$, which is experimentally feasible at the moment~\cite{Stock,Buckley}€, sensor response to shifts in $x$ is of the order of $10^{-3}(\text{nm})^{-1}$. Minimum detectable $x$ can be easily related to $\delta g_{\rm min}$ as $\delta x_{\rm min}=\frac{-1}{bg}\delta g_{\rm min}$. Figure \ref{fig4}(c) shows this noise equivalent $x$ is of the order of $10^{-3}\rm{(nm/\sqrt{Hz})}$. Young's modulus for GaAs is $E=85.5 \text{GPa}$~\cite{Sze}. Therefore, for the given lattice with a thickness of $t\simeq1\mu\text{m}$  the  stiffness of GaAs is $k=\frac{E}{t}\simeq 85.5 \frac{\text{kN}}{\text{m}}$. Minimum detectable force is shown in Fig. \ref{fig4}(d) and is of the order of $10^{-7}\rm N$ which compares rather well with the high resolution PhC force sensors~\cite{Ji,Yang} exploiting coherent light. However, these schemes use significantly larger input power while in our results not only is the pulse power (1 ph/pulse) low but also the average power ($10^{-10}$ W), which is defined by the emission rate of the current single photon sources ($\sim$ GHz), is also low. 

Importantly, fabricating the cavities with a smaller $\kappa$ does not affect the sensor resolution but shifts the optimum operating points at which the sensor behaves linearly (white dashed lines in Fig. \ref{fig4}) towards larger $x_{\rm bias}$.
\begin{figure}
\centering
\includegraphics[width=\linewidth]{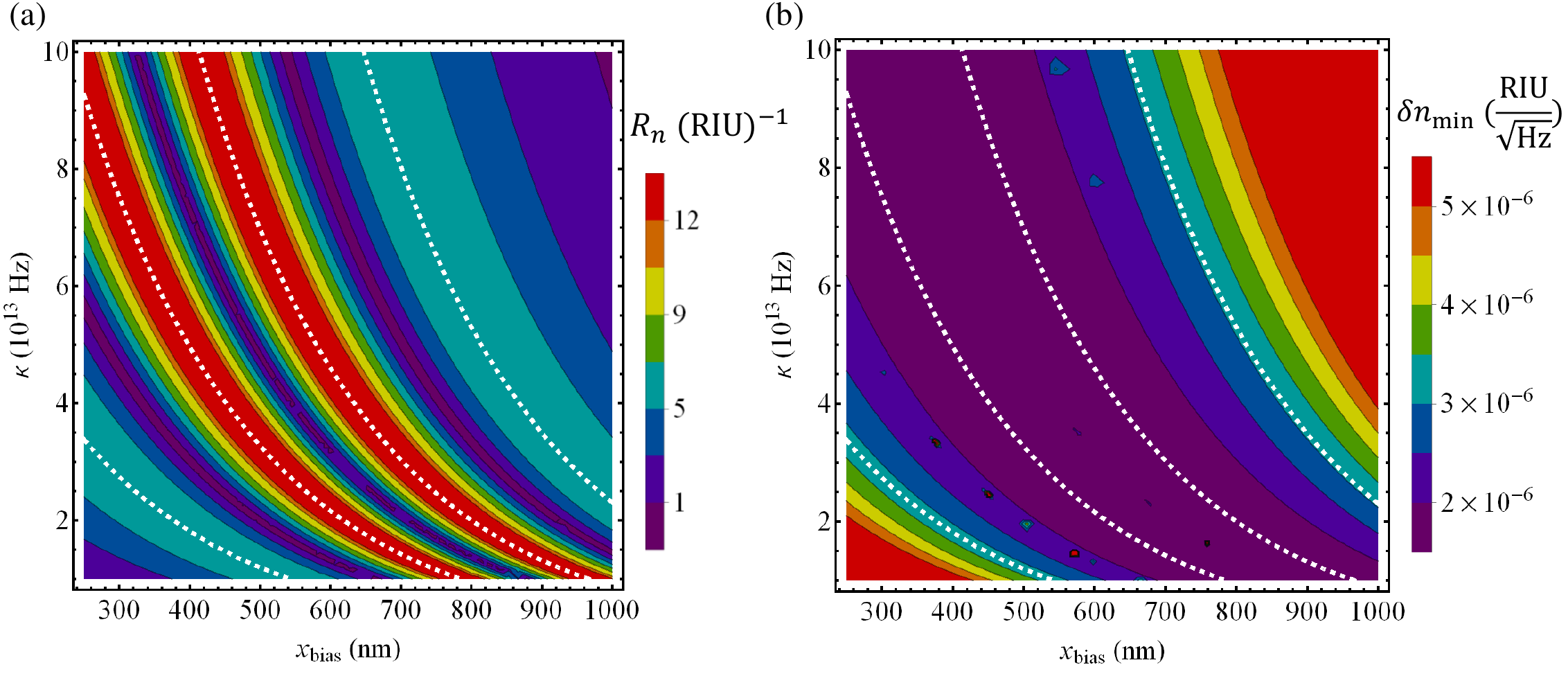}
\caption{Performance of HOM sensor as a refractive index sensor. This figure is a fabrication guide to building a HOM refractive index sensor with maximum performance. (a) Shows responsivity of the refractive index sensor for different operating points $x_{\rm bias}$ for an input photon bandwidth of $\gamma=1\rm GHz$. The white dashed lines show the bias points where sensor response changes linearly for very small changes in refractive index. Our theory predicts that responsivity does not depend on single photon band width $\gamma$. (b) Predicts that for $\gamma=1\rm GHz$ the  minimum detectable refractive index shift is of the order of $10^{-6}\rm RIU/\sqrt{Hz}$. }
\label{fig5}
\end{figure}

\subsection{HOM refractive index sensor}

To operate the system as a refractive index sensor we operate at a fixed $x_{\rm bias}$, so $g(n)=ae^{-bx_{\rm bias}+dn^2}$. System response to refractive index shift is $R_n(x_0,\kappa)=\vert\frac{dG^2(0)}{dn}\vert_{n=1}$ and the minimum detectable refractive index shift is calculated as $\delta n_{\rm min}=(\delta g_{\rm min}/2dgn)\vert_{n=1}$. Figure \ref{fig5}(a) is a fabrication guide for $\gamma=1\rm GHz$ to find the best operating points to achieve maximum responsivity together with linear response. Figure \ref{fig5}(b) predicts a resolution of the order of $10^{-6}$ refractive index unit (RIU) per $\rm \sqrt{Hz}$ for single photon bandwidth of $\gamma=1\rm GHz$. Up to the best of our knowledge the best resolution achieved in schemes~\cite{Wu1,Wu2} is of the order of $10^{-7}$RIU per $\rm \sqrt{Hz}$, however these use more input power.

\section{Conclusion}

In conclusion we have described a uniquely quantum protocol for a PhC sensor based on two coupled cavities. Our proposal uses single photon states, not coherent states, and operates on Hong-Ou-Mandel interference; a fourth order interference effect. The visibility for such a HOM proposal is dependent  on changes in the coupling between the cavities, which is in turn dependent on shifts in the cavity separation distance and/or refractive index of the medium in between the two cavities. Very small changes in such parameters result in modulation of the cavity coupling rate that can be observed by measuring the resulting change in the HOM dip. Our results predict minimum detectable values for refractive index and force changes, $10^{-6}$ RIU per $\rm \sqrt{Hz}$ and $10^{-7}$ N, respectively. 
This estimation is based on the parameters obtained form the current experimental implementations of coupled L3 PhC cavities in GaAs/AlGaAs~\cite{data} and is not specifically optimized for our sensor. Further development of PhC technology for the sensor presented here could offer significant improvements in the performance.  
Our results show that high sensitivity can be reached upon achieving high repetition rate single photon sources.

The advantages of the presented scheme are as follows. This scheme can be implemented on chip and fabricated in micro-scale dimensions. Moreover, unlike sensing approaches based on transmission spectrum of a L3 cavity coupled to a waveguide, this approach does not require spectral resolution that reduces the bandwidth. Additionally, this scheme can be a multi-purpose sensor. In this article, we have discussed force  (strain) and refractive index sensing. With minor modifications, it can be used for other targets such as local temperature, pressure and particle detection and analysis. 

A $\sqrt{2}$ improvement in estimation accuracy over schemes using coherent light and a $\sqrt{2}$ improvement in bandwidth over schemes using serial single photons can be achieved. The improvement of $\sqrt{2}$ in estimation accuracy may sound underwhelming. But a different accounting philosophy shows why this factor is important (see Appendix D for further details). Suppose we want our estimate of $g$ to have a mean square error of order $10^{-4}$ and we ask how many experiments, on average, we must perform to acheive this precision. Then coherent light  would require $50\times 10^6$ experiments; a serial single photon approach requires $24\times 10^6$ experiments; our HOM approach requires $12.5\times 10^6$ experiments. That is we have quartered the number of required experiments relative to the coherent light case and halved it relative to the serial single photon case. Due to this the reduction in samples the bandwith of our sensor relative to both cases is increased.

The disadvantage of this single-photon-based scheme compared to those using coherent light is the difficulty in building reliable single photon sources and detectors.

To summarise, the key point we are making in the paper is that if one has already made a commitment to single photonics in order to gain access to the quantum information processing that this provides, single-photon metrology can also be added to the suite of tools. HOM interference is the key phenomena that enables scalable quantum information processing in single photonics with linear optics. 

\section{Appendix A}
The  solutions to the quantum Langevin equations (\ref{stochastic_eqns}) are given by
\begin{eqnarray}
 a_1(t)&=&\sqrt{\kappa}\Big[A(t)\int_0^t dt^{\prime}\Big(C(t^\prime)a_{1,{\rm in}}(t^\prime)+D(t^\prime)a_{2,{\rm in}}(t^\prime)\Big)\nonumber \\ &&+B(t)\int_0^t dt^{\prime}\Big(D(t^\prime)a_{1,{\rm in}}(t^\prime)+C(t^\prime)a_{2,{\rm in}}(t^\prime)\Big)\Big],\nonumber \\
 a_2(t)&=&\sqrt{\kappa}\Big[B(t)\int_0^t dt^{\prime}\Big(C(t^\prime)a_{1,{\rm in}}(t^\prime)+D(t^\prime)a_{2,{\rm in}}(t^\prime)\Big)\nonumber \\&&+A(t)\int_0^t dt^{\prime}\Big(D(t^\prime)a_{1,{\rm in}}(t^\prime)+C(t^\prime)a_{2,{\rm in}}(t^\prime)\Big)\Big],
 \label{Langevin_sol}
\end{eqnarray}
where $A(t)=e^{-\kappa t/2}\cos (gt)$, $B(t)=-ie^{-\kappa t/2}\sin (gt)$, $C(t)=e^{\kappa t/2}\cos (gt)$ and $D(t)=ie^{\kappa t/2}\sin (gt)$. By using the above solutions for the cavity mode in the input-output relation (\ref{in-out}), we can analytically calculate the joint detection probability as

\begin{eqnarray}
G^{(2)}(\tau)=\dfrac{e^{-\dfrac{3}{2}\tau (\kappa+\gamma)}}{A}\left(Be^{\dfrac{3}{2}\tau (\kappa+\gamma)}+Ce^{-\dfrac{1}{2}\tau (3\kappa+\gamma)}\right. 
\left.+De^{\dfrac{1}{2}\tau (\kappa+3\gamma)}+Ee^{\tau (\kappa+\gamma)}\right),
\end{eqnarray}
where
\begin{equation}
A=(4g^2 + \kappa^2)^2\bigg(16 g^4 +(\gamma^2 -\kappa^2)^2+8 g^2 (\gamma^2 + \kappa^2)\bigg)^2,\nonumber
\end{equation}
\begin{eqnarray}
B&=& (4g^2 + (\gamma - \kappa)^2)^2 \bigg(256g^8 + \kappa^4 (\gamma+ \kappa)^4 
+8g^2 (\gamma^2 -2 \kappa^2) (16g^4 + \kappa^2 (\gamma+ \kappa)^2)\nonumber \\
&&+16g^4(\gamma^4 + 2\gamma^2 \kappa^2 +20 \gamma\kappa^3 + 22\kappa^4)\bigg),\nonumber
\end{eqnarray}
\begin{equation}
C=-32g^2\kappa^2(4 g^2 +\gamma^2 -\kappa^2)^2(4g^2 + \kappa^2)^2,\nonumber
\end{equation}
\begin{equation}
D = -32 g^2 \gamma^2 \kappa^2 F^2,\nonumber
\end{equation}
\begin{equation}
E = -64g^2\gamma\kappa^2(4g^2 +\gamma^2 -\kappa^2)(4g^2 +\kappa^2)F,\nonumber
\end{equation}
and
\begin{eqnarray}
F=\kappa(-12g^2 - \gamma^2 + \kappa^2)\cos(g\tau) 
+ 2 g (4 g^2 + \gamma^2 - 3 \kappa^2) \sin(g\tau).\nonumber
\end{eqnarray}

\section{Appendix B}

\begin{figure}[htbp!]
 \centering
\includegraphics[width=\linewidth]{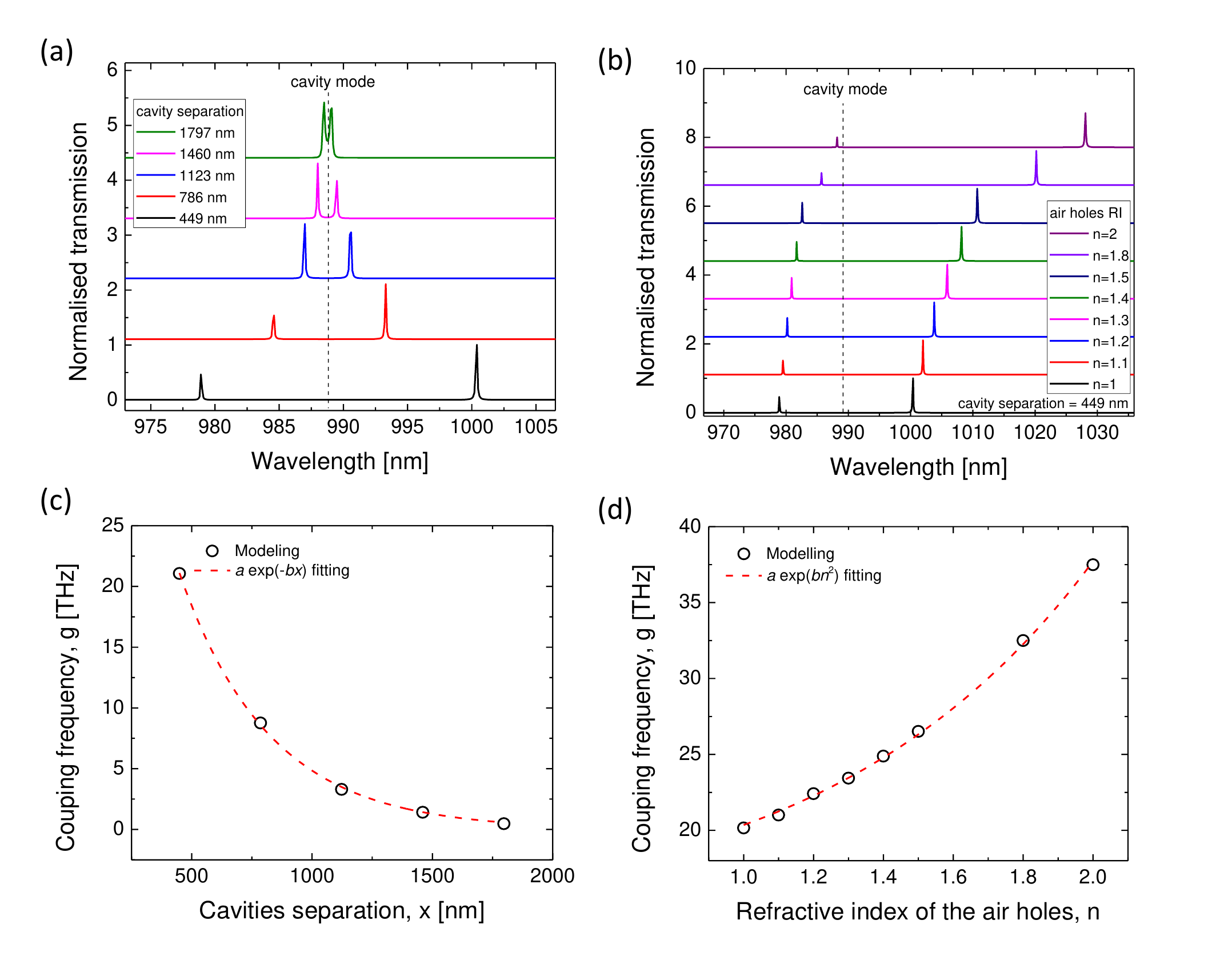}
\caption{Approximating the optical coupling between two GaAs cavities (L = 450 nm) placed in the middle of a distributed Bragg reflector (DBR) stack comprising GaAs (d=225 nm) and air (d=112 nm) pairs versus cavities separation (a,c) and air holes refractive index (b,d). (a) Normalised transmission of the stack showing the normal modes of the two cavities for different x. The dashed line corresponds to the unperturbed single cavity mode confined by the DBR stack. (b) Normal modes shown for x=449 nm (two air and one GaAs layers) and varying the refractive index of the air layers between the cavities. By increasing the air hole refractive index the mode separation increases which corresponds to stronger coupling between the cavities due to the decrease in the refractive index offset of the DBR layers. (c) coupling frequency calculated from (a) versus cavity separation. The dashed line corresponds to an exponential decay fitting. (d) coupling frequency as a function of air hole refractive index calculated from figure (b). The dashed curve corresponds to an exponential fitting of $ae^{bn^2}$. }
\label{figA1}
\end{figure}

To find the functionality of coupling strength $g$ with separation distance between the cavities and refractive index of the media in between the two cavities, we can use the analogy of the coupled cavities with a quantum double-well problem with a potential barrier in between, where we need to find the tunnelling rate $g$. Solving this problem shows the functionality of $g$ with the width of the barrier $x$ and the height of the barrier $V$ scales as $g\propto \exp\lbrace -bx-dV\rbrace$. According to citations~\cite{Marte,Markku}, which introduce the optical equivalence of the Scr\"odinger equation, $V\propto -n^2$. Therefore, we expect the coupling strength between the cavities to scale as $g\propto \exp\lbrace -bx+dn^2\rbrace$. To be more precise a one dimensional optical modelling simulation has been performed to find the functionality of coupling strength $g$ with $x$ and $n$, which suggest a very good fit can be achieved with the above given functionality. The simulations are done by the transfer matrix method described in \cite{Pettersson}. The results are shown in Fig. \ref{figA1}.

\newpage
\section{Appendix C}

\begin{figure}[htbp!]
\centering
\includegraphics[width=0.7\linewidth]{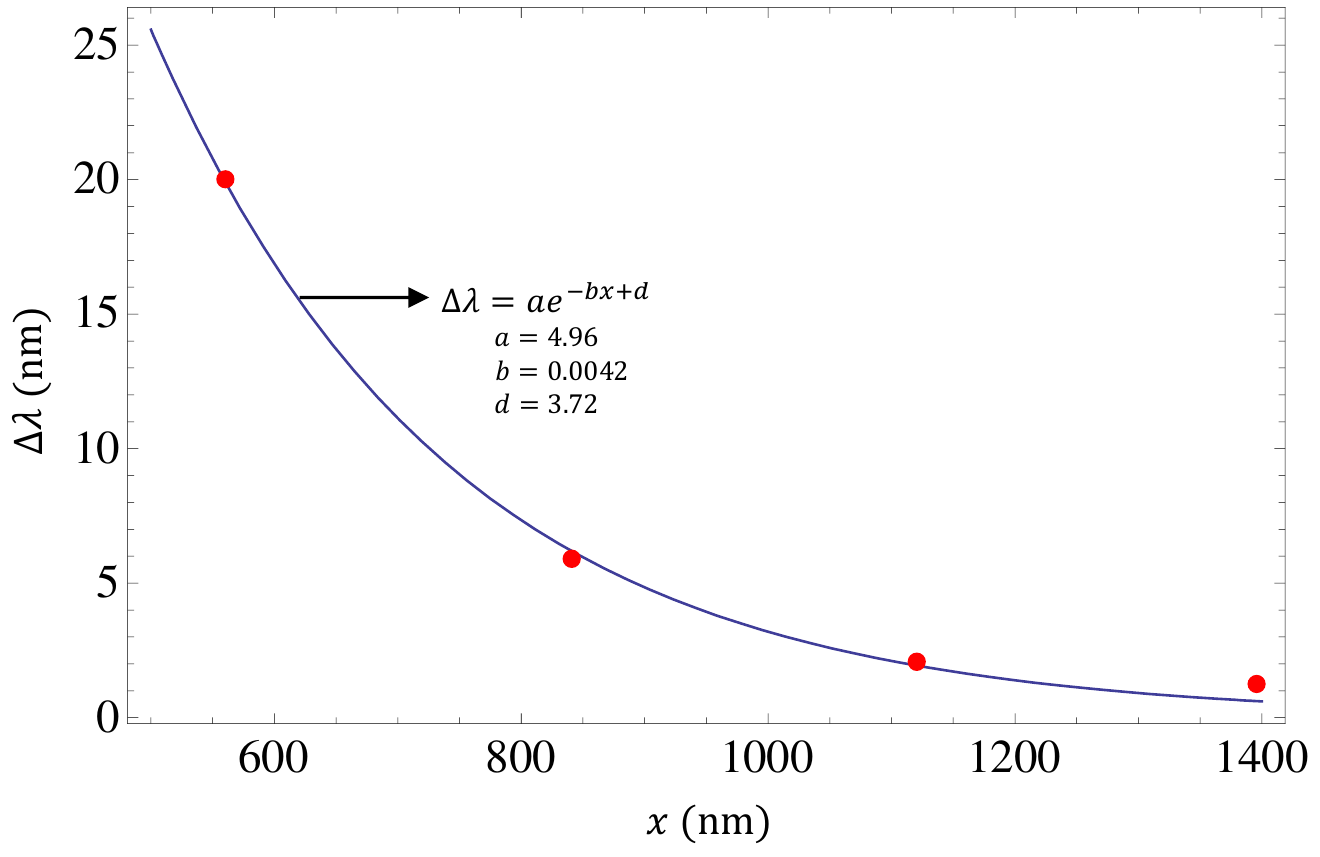}
\caption{Functionality estimation of $g$ versus $x$ and $n$. Fit on experimental data (red points) given in figure 2 of citation \cite{data} to find the coefficient a, b and d in functionality of $g$ versus $x$ and $n$ that we found of the form $g(x,n)=\pi c \Delta\lambda/\lambda^2= \frac{\pi c}{\lambda^2}ae^{-bx+dn^2}$ where $\lambda=1000 \rm nm$.}
\label{figA2}
\end{figure}

We choose experimental data for a PhC lattice made of GaAs/AlGaAs given in~\cite{data} as an example to find the functionality of coupling strength $g$ in terms of cavities separation $x$ and the refractive index of the dielectric material $n$. Figure \ref{figA2} shows the numbers for the best found fitted function.

\section{Appendix D}

Here we back up the claims made in the conclusion, further details will be available in a future publication. Here we assume a mean flux greater than two photons may damage a sample. We model our coupled cavity sensor as an effective beam splitter between the input modes such that the beam splitter reflectivity $g$ is a function of the parameter to be estimated, $x$, i.e $g(x)$. In the case of very small changes in $x$, this functional relationship can be linearized about an operating point $x_0$ so that changes in $\theta$ are linearly proportional to changes in $g$. 

Now we ask what is the ultimate limit imposed by quantum theory on the precision of our estimate $g_{\rm est}$ of $g$. This, of course depends on the state used to probe the beam-splitter. If the mean squared error (MSE) i.e. $\mathbb{E}[(g -g_{\rm est})^2]$ quantifies the performance of our estimation scheme then the quantum Cramer-Rao bound provides a method to answer this question. The quantum Cramer-Rao bound states that asymptotically the precision of an unbiased estimation scheme is bounded below by $1/\sqrt{N I(g)}$ where $N$ is the number of experimental trials and $I(g)$ is the quantum Fisher information with respect to the input state.

If the input state is a coherent state then the quantum Fisher information is  generally $I_\alpha(g) = |\alpha|^4$ where $|\alpha|^2$ is the mean photon number at the input. So for the cases of interest $I_\alpha(g) = 1$ and $I_\alpha(g) = 4$. If the input state is a Fock state then the quantum Fisher information is \cite{loss}: $I_F(g) = 4$ (input photon number $=1$), $I_F(g) = 16$ (input photon number $=2$). The numbers quoted in the main text can be arrived at by setting MSE $=10^{-4}$ and solving for $N$.

Loss is easily included using the standard beam splitter model discussed in section 2. Coherent states are not entangled on beam splitters so tracing out the vacuum modes leaves the signal mode in a coherent state with amplitude reduced by $\alpha\rightarrow t\alpha$, where $|t|^2=T$ is the probability for a photon {\em not} to be lost. The Fisher information~\cite{Gerard_Fisher} is then simply rescaled to reflect the loss of amplitude. Single photon product states at the input to a beam splitter do become entangled at the output. However, we can trace out the vacuum modes to give a mixed input state for the case of $n=2$ in a HOM experiment of the form 
\begin{eqnarray}
\rho_{\rm in} & = & T_1T_2 |1\rangle_1\langle 1|\otimes|1\rangle_2\langle 1|+T_1R_2 |1\rangle_1\langle 1|\otimes|0\rangle_2\langle 0|\\\nonumber
&&+R_1T_2 |0\rangle_1\langle 0|\otimes|1\rangle_2\langle 1|+R_1R_2 |0\rangle_1\langle 0|\otimes|0\rangle_2\langle 0\vert.
\end{eqnarray}
The Fisher information for a mixed state is more difficult to calculate as it involves the logarithmic derivative. However, because the mixed state above is so simple it is easy to see that, conditioned on detecting two photons for correct heralded operation, the Fisher information is rescaled by $T_1^2T_2^2$. 

\section*{Acknowledgments}
We acknowledge the support of the Australian Research Council Centre of Excellence for Engineered Quantum Systems, CE110001013. SBE and AA were funded by the University of Queensland International Scholarship. JC was supported in part by National Science Foundation Grant Nos. PHY-1212445 and PHY-1314763. Authors thank Martin Ringbauer, Michael Vanner and Devon Biggerstaff for helpful discussions.


\end{document}